\documentclass{article}
\usepackage{float}
\usepackage{graphicx} 
\usepackage{amsmath} 
\usepackage{parskip} 
\usepackage{dsfont} 
\usepackage{geometry}  
\usepackage{graphics} 
\usepackage{color}
\usepackage{hyperref}
\usepackage{setspace}
\usepackage[linesnumbered,ruled,vlined]{algorithm2e}
\newgeometry{vmargin={15mm}, hmargin={30mm,30mm}}   

\title{\bf Bivariate inverse Gaussian degradation processes with shared random effects and an application to fatigue cracks}
\author{{Yuvraj Dutta}${}^{1}$, {Sandip Barui}${}^{1}$, {Debanjan Mitra${}^{2\footnote{Corresponding author. E-mail address: debanjan.mitra@iimu.ac.in \newline \indent}}$, Narayanaswamy Balakrishnan${}^{3}$}  \\\\
${}^{1}$Interdisciplinary Statistical Research Unit, \\ Indian Statistical Institute, Kolkata, 700108, West Bengal, India\\ [1ex]
${}^{2}$ Operations Management, Quantitative Methods, and Information Systems Area, \\
Indian Institute of Management Udaipur, Udaipur, 313001, Rajasthan, India \\ [1ex]
${}^{3}$Department of Mathematics and Statistics,\\ 
McMaster University, 1280 Main Street West, Hamilton, Ontario L8S 4K1, Canada\\
}
\date{}

\begin{document}

\maketitle

\begin{abstract}
The inverse Gaussian (IG) process is a widely used model for univariate degradation data. For bivariate degradation data involving two performance characteristics (PCs), dependence is often introduced through an unobserved shared frailty factor combined with IG processes. Previous studies typically assume a specific frailty distribution, such as normal or gamma, although such choices are difficult to justify because the frailty is unobserved. This paper proposes a general IG–GG framework for modeling bivariate degradation data with dependent PCs. Each degradation process is modeled using an IG process, while the shared frailty follows the generalized gamma (GG) family, which includes exponential, gamma, Weibull, and lognormal distributions as special cases. The proposed framework allows flexible selection of an appropriate frailty distribution within the GG family, leading to improved model fitting. Convenient parameter estimation procedures are developed and evaluated through simulation studies, demonstrating satisfactory performance. The proposed model is applied to fatigue-crack data and compared with several existing frailty-based and copula-based models. Results show that the IG–GG model provides a superior fit. System reliability estimation under the IG–GG framework is also discussed.
\end{abstract}

\vspace{0.1 in}



\setstretch{1.5}

\section{Introduction} \label{sec:intro}
Monitoring the continuous degradation of relevant performance characteristics (PCs) of engineered systems over time provides valuable information about system lifetimes, particularly when failure data are scarce and expensive to obtain. Consequently, a substantial body of research has emerged on degradation modeling, encompassing a variety of models and methodologies for analyzing PC-level data collected from complex systems. A primary goal of degradation modeling is to infer system lifetime characteristics from observed degradation data. Broadly, degradation modeling approaches may be categorized into path models and stochastic process models \cite{Meeker-book}; for a comprehensive overview of current developments in this area, see the recent volume by Chen et al.~\cite{Tony-book}. Among stochastic process models for degradation analysis, several random processes have been extensively studied, with the Wiener process (\cite{Whitmore}, \cite{Wang-Weiner}, \cite{ZHANG}), gamma process (\cite{Abdel-Hameed}, \cite{Lawless-Crowder}), and inverse Gaussian (IG) process (\cite{Wang-Xu}, \cite{Ye-Chen}) being the most prominent. Each of these processes possesses distinct characteristics that make it suitable for modeling particular types of degradation behavior.

When degradation in two or more PCs of a complex system is examined to gain a deeper understanding of system lifetime characteristics, an additional challenge arises because the PCs are often inherently dependent. Examples include positioning accuracy and output power in heavy machine tools~\cite{Peng}, braking torque and response time in the evaluation of demagnetization of permanent magnet brakes~\cite{Zhuang}, and rubidium consumption and intensity in rubidium discharge lamps~\cite{Pan-Bala}. To model degradation in multiple PCs, copula-based approaches have been widely adopted. Under this framework, the marginal degradation processes of the PCs are first modeled separately using appropriate stochastic processes, after which the resulting random processes are linked through a copula function. Various copula families have been employed for this purpose, with one of the earliest contributions being the work of Sari et al.~\cite{Sari}. Subsequently, numerous studies have adopted this strategy. For instance, Pan et al.~\cite{Pan2013} modeled bivariate degradation data by first describing the marginal degradation using Wiener processes and then coupling them through Gaussian, Frank, Gumbel, and Clayton copulas. Additional discussions and developments can be found in Fang and Pan~\cite{Fang-Pan}, Fang et al.~\cite{Fang2020}, and the references cited therein.

More recently, frailty-based approaches have been investigated for modeling dependent degradation data. In this framework, the marginal degradation processes corresponding to the PCs are represented by stochastic processes and are linked through an unobserved common random component known as the frailty. Compared with copula-based methods, frailty-based models offer certain advantages, including a more intuitive conceptual interpretation and greater convenience in estimating important reliability measures. Besides capturing dependence among components, the frailty term also accounts for unit-to-unit heterogeneity in system behavior. Xu et al.~\cite{Xu2018} employed a normally distributed shared frailty in conjunction with Wiener process degradation models. Song and Cui~\cite{Song-Cui} considered gamma process marginals along with a gamma frailty model. Barui et al.~\cite{Barui} further generalized this framework for bivariate gamma degradation processes by introducing a generalized gamma frailty distribution for the latent shared random effect. For a comprehensive review of multivariate degradation modeling methods, including frailty- and copula-based approaches, readers are referred to the recent and thorough survey by Yi et al.~\cite{Yi-Review}.

Among stochastic degradation models, the IG process has attracted considerable attention in recent years. In particular, random-effects versions of the IG process have been actively studied because the inclusion of random effects substantially enhances the flexibility of the model for degradation analysis; see, for example, Shahraki et al.~\cite{Shahraki} and the references therein. Wang and Xu~\cite{Wang-Xu} and Ye and Chen~\cite{Ye-Chen} strongly advocated the use of the IG process for degradation modeling, demonstrating that it can be more flexible than gamma process models in certain situations. The IG process is especially suitable for incorporating random effects to account for unit-to-unit variability in univariate degradation data. Wang and Xu~\cite{Wang-Xu} incorporated a gamma-distributed random effect into their IG process model. Ye and Chen~\cite{Ye-Chen} extended this framework by introducing random effects into both the drift and volatility parameters of the IG process. Hao et al.~\cite{Hao} further proposed an extended IG process model involving a skew-normal random effect for degradation analysis. Bayesian inference for IG process degradation models was discussed by Peng et al.~\cite{Peng-Bayesian}, while Wang et al.~\cite{Wang-ADT} applied the IG process in accelerated degradation testing.

The IG process has also been employed in the analysis of bivariate and multivariate degradation data. Peng et al.~\cite{Peng} developed a copula-based model for bivariate degradation data using IG process marginals. For a detailed treatment of copula-based modeling of dependent degradation data with various marginal processes, including the IG process, readers may consult Fang and Pan~\cite{Fang-Pan}. Frailty-based methodologies have likewise been explored in conjunction with IG process models. Recently, Fang et al.~\cite{Fang2022} considered multivariate degradation data where each marginal degradation process follows an IG process and the shared random effect is modeled through a multivariate normal distribution. In their framework, dependence among components is characterized through the covariance structure of the multivariate normal distribution. Duan et al.~\cite{Duan} investigated Bayesian inference for bivariate degradation data under an IG process model in which the drift and volatility parameters are associated with truncated normal and gamma random effects, respectively. More recently, Zhuang et al.~\cite{Zhuang} introduced a multivariate reparameterized IG process for multivariate degradation modeling.

An important consideration in frailty-based modeling is the selection of an appropriate distribution for the shared random effect, namely the frailty distribution. This issue is particularly significant because the frailty is an unobserved latent variable that influences the degradation behavior of the components and therefore lacks a direct empirical counterpart in the observed data. Although frailty-based models are highly flexible and powerful, determining whether a specific frailty distribution adequately represents a given dataset remains a challenging problem, especially because the shared random effect itself cannot be directly observed. The generalized gamma frailty model provides a useful solution in this context for bivariate degradation data \cite{Barui}. A key advantage of employing a broad family of frailty distributions is that, for a given dataset, one can identify and select an appropriate frailty distribution from within the family rather than imposing a predetermined choice. This flexibility can lead to a more accurate representation of the latent shared effect than assuming a fixed frailty distribution. Existing studies on bivariate and multivariate degradation modeling with IG processes have generally relied on specific frailty distributions, thereby limiting the ability to choose the most suitable frailty model based on the observed data.

In this paper, we propose a flexible frailty-based framework for modeling bivariate degradation data. The marginal degradation processes associated with the two PCs are modeled using inverse Gaussian processes, while the shared latent random effect is assumed to follow the generalized gamma (GG) family of frailty distributions. This framework preserves the desirable properties of the IG process for marginal degradation modeling while simultaneously allowing the most appropriate frailty distribution within the GG family to be selected for a given bivariate dataset. Within this setting, we also address the estimation of system reliability for systems with two degrading components while accounting for the common frailty effect. Through analysis of a bivariate fatigue-crack degradation dataset, we demonstrate that the proposed model provides a superior fit compared with several widely used copula-based alternatives.

The remainder of the paper is organized as follows. Section 2 presents the necessary background on potentially dependent degradation processes of performance characteristics (PCs), whose marginal behaviors are modeled using inverse Gaussian distributions, along with the key properties of the generalized gamma shared frailty distribution. Section 3 introduces the proposed IG–GG modeling framework and provides the motivation for its use in the analysis of bivariate degradation data. Parameter estimation procedures and the computation of standard errors following model fitting are described in Section 4. Section 5 presents a comprehensive simulation study conducted under various parameter settings to evaluate the performance of the proposed methods. In Section 6, the proposed model is applied to a real-world fatigue-crack dataset, and its performance is compared with several well-established copula-based models. Section 7 discusses the estimation of system reliability under the proposed framework. Finally, Section 8 concludes the paper with a summary of the main findings and some closing remarks.

\section{Background} \label{sec:background}
\subsection{Bivariate degradation data} \label{subsec:data-struc}
Consider \textit{n} units under test, for observing degradation of two PCs. For each unit, degradation of the two PCs are measured at time points $t_{ij}$, $j=1,\dots,m_i$, $i=1,\dots,n$. For each unit, both PC's are measured at each measurement time. 

Let $x_k(t_{ij})$ denote the measured degradation of the $k$-th PC for the $i$-th unit at time $t_{ij}$, $k$=1,2. Without loss of generality, the observed degradation for both PCs at the starting point are considered to be zero, i.e., we set $t_{i0}=0$, $x_k(0)=0$, $\forall$ $i=1,\dots,n$, $k$=1,2. 

\subsection{The inverse Gaussian process} \label{subsec:IG}
The inverse Gaussian (IG) process, denoted $\{Y(t)$$, t \geq 0\}$, is a random process satisfying the following properties: 
\begin{itemize}
    \item $Y(t)$ has independent increments, i.e., $Y(s)-Y(t)$ and $Y(u)-Y(v)$ are independent $\forall$ $ s > t \geq v > v \geq 0$;  
    \item $Y(t) - Y(s)$ follows $ IG(\mu(\Lambda(t;b)-\Lambda(s;b)),\eta(\Lambda(t;b)-\Lambda(s;b))^2)$, $\forall$ $t>s \geq 0$.
\end{itemize}
Here, $\Lambda(t;b)$ is a monotonic function to transform the time scale if non-linearity exists in the degradation process. The probability density function (PDF) of $Y(t)$ that follows $IG(a,b)$ is
\begin{equation}
f_{IG}(y;a,b)=\sqrt{\frac{b}{2 \pi y^3}}e^{-\frac{b(y-a)^3}{2a^2y}} \mathds{1}(y>0), \label{eq:IG}
\end{equation}
where $a>0$ is the mean parameter and $b>0$ is the shape parameter. The mean and variance of $Y(t)$ are given by
\[
E[Y(t)] = a, \quad Var[Y(t)] = a^3/b. 
\]
For further details on the IG process, refer to Chhikara and Folks~\cite{Chhikara-IG}, and Sheshadri~\cite{Sheshadri-IG}.  

\subsection{The generalized gamma distribution} \label{subsec:GG}
The GG distribution is a family or class of distributions, containing some of the commonly used probability distributions as particular members of the family. The PDF of a random variable $T$ that follows $GG(q, \sigma, \alpha)$ distribution is given by 
\begin{equation}
f_{gg}(t ; q,\sigma,\alpha)=\begin{cases} 
     \frac{|q|(q^{-2})^{(q^{-2})}(\alpha t)^{(q^{-2})(\frac{q}{\sigma})}}{\Gamma(q^{-2})\sigma t} e^{-(q^{-2})(\alpha t)^{(\frac{q}{\sigma})}}, & q\neq 0; \\
      \frac{1}{\sqrt{2\pi}\sigma t}e^{\frac{-(log(\alpha t))^2}{2 \sigma^2}}, & q=0,
   \end{cases} \label{eq:GG}
\end{equation}   
with $\alpha$ as the scale parameter, and $\sigma$ and $q$ as the shape parameters, $\alpha>0$, $\sigma>0$, $-\infty<q<\infty$. For more details on the GG family of distributions, see Balakrishnan and Peng~\cite{Bala-Peng} and the references therein. 

The GG family derives its flexibility as a parametric model due to the presence of the two shape parameters, different values of which lead to different special members of the family. The exponential, gamma, and Weibull distributions are all special members of the GG family, while the lognormal distribution is its limiting case. In particular, the specific values of the shape parameters for the various special members of the GG family are given in Table \ref{tab:GG}. 
\begin{table}
\caption{Special members of the GG family of distributions}
\begin{center}
\begin{tabular}{|c|c|} 
 \hline
 Conditions & Distribution  \\ 
 \hline
 $q=\sigma=1$ & Exponential  \\ 
 $q=1$  & Weibull  \\ 
 $\frac{q}{\sigma}=1$  & Gamma  \\ 
 $q=0$ or $q \xrightarrow{} 0$  & Lognormal  \\ 
\hline
\end{tabular} \label{tab:GG}
\end{center}
\end{table}
The mean and variance of a random variable following the GG distribution are as follows: \\
Mean = $\frac{\Gamma(q^{-2}+\frac{\sigma}{q})}{\Gamma(q^{-2})(q^{-2})^{\frac{\sigma}{q}}\alpha}$, \\
Variance = $\frac{1}{\Gamma(q^{-2})(q^{-2})^{\frac{2\sigma}{q}}\alpha^2}[\Gamma(q^{-2}+\frac{2\sigma}{q})-\frac{[\Gamma(q^{-2}+\frac{\sigma}{q})]^2}{\Gamma(q^{-2})}]$. 

DiCiccio~\cite{Diciccio} and Lawless~\cite{Lawless-GG} are some of the first authors to discuss inference for the GG distribution; see also Noufaily and Jones~\cite{Nofaily-GG} for recent developments with regard to estimation for this distribution. 

\section{Frailty-based IG process model for bivariate degradation data} \label{sec:model}
Let $X_k(t)$ denote the degradation process corresponding to the $k$-th PC, $k=1,2$. Suppose $\lambda$ denotes the shared random effect that influences the two degradation processes corresponding to the PCs. Conditional on the shared frailty $\lambda$, it is assumed that the two degradation processes are independent and marginally follow the IG process. The model further assumes that the shared frailty term $\lambda$ follows the GG distribution with its PDF as in Eq.\eqref{eq:GG}. 

The hierarchical model structure is thus as follows: 
\begin{align}
	X_1(t)|\lambda & \sim IG(\lambda\mu_1\Lambda(t;b_1),\eta_1(\Lambda(t;b_1))^2), \nonumber \\
	X_2(t)|\lambda & \sim IG(\lambda\mu_2\Lambda(t;b_2),\eta_2(\Lambda(t;b_2))^2), \nonumber \\
	\lambda & \sim GG(q, \sigma, \alpha), \label{eq:model-structure}
\end{align} 
where $\Lambda(t;b_k)$ is a monotonic function of $t$, $k=1,2$. The PDF of $X_k(t)$, the degradation process corresponding to the PCs, conditional on the frailty $\lambda$, is given by
\begin{equation} 
f_{X_k(t)}(x_k(t))=\sqrt{\frac{\eta_k}{2\pi (x_k(t))^3}}\Lambda(t;b_k)e^{-\frac{\eta_k(\Lambda(t;b_k))^2(x_k(t)-\lambda\mu_k\Lambda(t;b_k))^2}{2\lambda^2\mu_k^2(\Lambda(t;b_k))^2(x_k(t))}}. \label{eq:IG-cond}
\end{equation}

Note that, by assuming the shared frailty term to follow the GG distribution, we actually include a class of distributions for the frailty term, and avoid specifying a particular distribution for the unobserved random effect. For modeling any bivariate degradation data, the proposal is to fit all the special members of the GG family, along with the GG distribution itself, to the data. Then, depending on a suitable model selection criterion such as the Akaike's information criterion (AIC) or Bayesian information criterion (BIC), the most suitable model within the GG family may be chosen for the observed bivariate degradation data. This approach is particularly helpful with frailty-based models, as there is no direct way for conducting statistical hypothesis tests for identifying the frailty distribution for a concerned data, frailty being an unobserved quantity. 

\begin{figure}[ht]
	\centering
	\includegraphics[width=\textwidth]{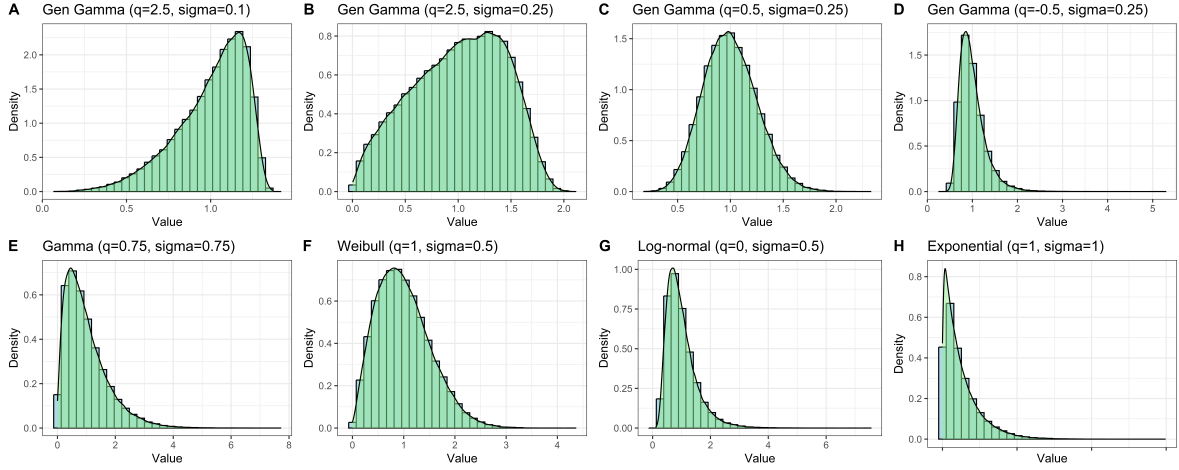}
	\caption{Density of GG distribution and the special members of the GG family}
	\label{fig:GG family}
\end{figure}

This modeling approach helps one to detect the suitable distributional form of the frailty that induces dependence among the two degrading components, which would have otherwise been missed if a specific frailty distribution such as gamma or Weibull was used instead of the general family of generalized gamma. The GG distribution takes various shapes - both as special members of the family and as the distribution itself - for various choices of values of its parameters. The GG frailty in bivariate degradation model is thus capable of capturing a wide variety of dependence structures between the degrading components. In Figure \ref{fig:GG family}, the density functions of the GG distribution and its special cases are presented for different choices of the shape parameters $q$ and $\sigma$. 

We demonstrate this point with the help of a case study based on a simulated bivariate degradation data in the following sub-section. Note that the fitting methods that are used to fit the proposed model to this simulated bivariate degradation data will be discussed in Section \ref{sec:fitting} in detail.

\subsection{Motivating example} \label{subsec:motivation}
To provide a motivating example for considering the proposed IG-GG modeling framework for bivariate degradation data, we present here a simulated data. The data is simulated from a bivariate IG process with shared random effect following the GG distribution with parameter values as follows: $b_1 = 0.80$, $\eta_1 = 1.50$, $\mu_1 = 2.50$, $b_2 = 1.20$, $\eta_2 = 1.30$, $\mu_2 = 3.50$, $q = 2.50$, $\sigma = 0.30$, and $\alpha = 0.69$. Note that the choice of value for the parameter $\alpha$ is obtained by setting expectation of the shared frailty distribution to be 1. In the simulated data, the number of units is taken as $n = 100$, and for each unit, the degradation of the two PCs are measured at $m = 20$ time points, where $t_1 = 2$, $t_2 = 4$, $t_3 = 6$,...,$t_{20} = 40$, with $t_0 = 0$.

To this data, we fit the proposed model described in Section \ref{sec:model}. This results in the fitting of five different models, with the marginal degradation modeled by the inverse Gaussian process, and the frailty modeled by the members of the GG family of distributions. We denote these models by $\mathcal{M}_1$ for exponential frailty, $\mathcal{M}_2$ for gamma frailty, $\mathcal{M}_3$ for Weibull frailty, $\mathcal{M}_4$ for lognormal frailty, and $\mathcal{M}_5$ for generalized gamma frailty. Although the model fitting methods will be discussed in the next section in detail, we present the results of model fits in Table \ref{tab:simulated-data-estimates} here. Among the fitted models, the one with generalized gamma frailty has the lowest AIC, indicating that this is the best model for the observed bivariate degradation data. The AIC values corresponding to different fitted models are given in Table \ref{tab:simulated-data-AIC}.  

\begin{table}
	\caption{Estimates of parameters for different models in the IG-GG modeling framework; within parentheses are the corresponding standard errors of the parameter estimates}
	\begin{center}
		\begin{tabular}{|c|c|c|c|c|c|c|c|c|c|} 
			\hline
			Model & $b_1$ & $\eta_1$ & $\mu_1$ & $b_2$ & $\eta_2$ & $\mu_2$ & $\sigma$ & $q$ & $\alpha$\\ 
			\hline
			$\mathcal{M}_1$ & 0.813 & 1.954 & 2.970 & 1.220 & 2.002 & 3.665 & 1.000 & 1.000 & 1.000  \\ 
			& (0.026) & (0.461) & (0.415) & (0.032) & (0.518) & (0.441) & (0.000) & (0.000) & (0.000) \\
			$\mathcal{M}_2$ & 0.807 & 1.225 & 2.418 & 1.175 & 1.053 & 3.916 &	0.569 & 0.569 & 1.000 \\ 
			& (0.021) & (0.344) & (0.257) & (0.037) & (0.298) & (0.470) & (0.262) & (0.262) & (0.000)\\
			$\mathcal{M}_3$ & 0.802 & 1.463 & 2.488 & 1.199 & 1.344 & 3.532 & 0.731 & 1.000 & 0.920\\ 
			& (0.016) & (0.250)	& (0.234) & (0.031) & (0.349) & (0.294) & (0.114) & (0.000) & (0.020) \\
			$\mathcal{M}_4$ & 0.815 & 0.961 & 2.376 & 1.151 & 0.851 & 4.261 & 0.396 & 0.000 & 1.103	\\
			& (0.018) & (0.268) & (0.251) & (0.023) & (0.211) & (0.385) & (0.173) & (0.173) & (0.117)\\  
			$\mathcal{M}_5$ & 0.797 & 1.422 & 2.507 & 1.186 & 1.210 & 3.601 & 0.491 & 2.634 & 0.602	\\ 
			& (0.013) & (0.175) & (0.208) & (0.020) & (0.146) & (0.228) & (0.069) & (0.247) & (0.027) \\
			\hline
		\end{tabular} \label{tab:simulated-data-estimates}
	\end{center}
\end{table}

\begin{table}
	\caption{Log-likelihood, AIC, and BIC values for different models}
	\begin{center}
		\begin{tabular}{|c|c|c|c|} 
			\hline
			Model & Log-likelihood value & AIC value & BIC value\\ 
			\hline
			$\mathcal{M}_1$ & -10702.890 & 21417.780 & 21433.411 \\
			$\mathcal{M}_2$ & -10540.310 & 21094.620 & 21112.856 \\
			$\mathcal{M}_3$ & -10500.690 & 21015.380 & 21033.616 \\
			$\mathcal{M}_4$ & -10621.080 & 21256.160 & 21274.396 \\
			$\mathcal{M}_5$ & {\bf -10494.020} & {\bf 21004.040} & {\bf 21024.881} \\ 
			\hline
		\end{tabular} \label{tab:simulated-data-AIC}
	\end{center}
\end{table}

Clearly, the model with the generalized gamma as the frailty distribution turns out to be the best model for this data, as it has the lowest AIC and BIC values compared to the others. This shows the risk of using any specific distribution for modeling the frailty, as discussed in Balakrishnan and Peng~\cite{Bala-Peng}. If gamma or lognormal frailty was used for this data, the fit would have been sub-optimal. Only when one has the flexibility of choosing the suitable frailty distribution from available options, the fit is expected to be good. This demonstrates the advantage of using the generalized gamma class of distributions as frailty, and acts as the main motivation for developing a dependent degradation model based on IG process combined with generalized gamma frailty.

\section{Model fitting methods} \label{sec:fitting}
\subsection{Direct optimization of likelihood function} \label{subsec:direct}
Define
\[z_k(t_{ij})=x_k(t_{ij})-x_k(t_{i(j-1)}),\]
\[\omega(t_{ij};b_k)=\Lambda(t_{ij};b_k)-\Lambda(t_{i(j-1);b_k}).\]
Recall that, by IG process property mentioned in Section 2.2, we have 
$$Z_x(t_{ij}) \sim IG(\lambda_i\mu_k\omega(t_{ij};b_k),\eta_k(\omega(t_{ij};b_k))^2).$$

For observed bivariate degradation data with the structure described in Section 2.1, the likelihood function, conditional on the shared frailty term $\lambda$, is given by
\begin{equation}
L_{conditional}(\boldsymbol \theta|Data,\boldsymbol \lambda)=\prod_{i=1}^n\prod_{k=1}^2\prod_{j=1}^{m_i}\sqrt{\frac{\eta_k}{2\pi (z_k(t_{ij}))^3}} \times \omega(t_{ij};b_k)\\ \times e^{-\frac{\eta_k(z_k(t_{ij})-\lambda_i\mu_k\omega(t_{ij};b_k))^2}{2\lambda_i^2\mu_k^2(z_k(t_{ij}))}}, \label{eq:lik-cond}
\end{equation}
where $\boldsymbol \theta=(\mu_1,\mu_2,\eta_1,\eta_2,b_1,b_2)$ is the vector containing the model parameters of the marginal degradation processes, and $\boldsymbol \lambda=(\lambda_1,\dots,\lambda_n)$ is the vector of shared random effects, with $\lambda_i$ affecting the two degradation processes corresponding to the PCs for the $i$-th unit, $i=1,\dots,n$.

The marginal or unconditional likelihood function is obtained from \eqref{eq:lik-cond}, when the shared frailty term is integrated out. Thus, using the PDF of the GG distribution from \eqref{eq:GG} in \eqref{eq:lik-cond}, the marginal likelihood function is obtained as
\begin{equation} \label{eq:lik-marg}
\begin{split}
L_{marginal}(\boldsymbol \theta, \boldsymbol \phi|Data) 
&= \prod_{i=1}^n\prod_{k=1}^2\prod_{j=1}^{m_i}\sqrt{\frac{\eta_k}{2\pi (z_k(t_{ij}))^3}} \times \omega(t_{ij};b_k) \\
&\times \int_{0}^{\infty}e^{-\frac{\eta_k(z_k(t_{ij})-\lambda_i\mu_k\omega(t_{ij};b_k))^2}{2\lambda_i^2\mu_k^2(z_k(t_{ij}))}}f_{gg}(\lambda_i;q,\sigma,\alpha)d\lambda_i, 
\end{split}
\end{equation}
with $\alpha>0$, $\sigma>0$, $\infty < q < \infty$, $\eta_1,\eta_2>0$, $\mu_1, \mu_2 > 0$, $b_1, b_2>0$, for $k=1,2$. Here, $\boldsymbol \phi=(q,\sigma,\alpha)$ is the vector containing the parameters of the GG distribution which is used to model the shared frailty term. 

It is important to mention here that, to ensure identifiability of the model parameters, we must set the mean of the distribution of the shared frailty to be 1 while estimating the parameters. This is required in order to distinctly estimate the parameters of the random effect with respect to the parameters of the degradation process. This requirement to ensure identifiability of the parameters is one of the fundamental issues in frailty modeling, and one can refer to the book by Hougaard~\cite{Hougaard} for detailed discussions regarding this. Of course, setting the frailty mean as 1 would imply reduction in the number of parameters to be estimated; for the GG distribution, this condition implies 
\begin{equation}
\alpha= \frac{\Gamma(q^{-2}+\frac{\sigma}{q})}{\Gamma(q^{-2})(q^{-2})^{\frac{\sigma}{q}}}. \label{eq:identifiability}
\end{equation}

Evidently, the marginal likelihood in \eqref{eq:lik-marg} does not have a closed-form expression. Maximum likelihood estimates (MLEs) $\widehat{\boldsymbol \theta}$ and $\widehat{\boldsymbol \phi}$ of the model parameters are obtained by numerically optimizing \eqref{eq:lik-marg} with respect to $\boldsymbol \theta$ and $\boldsymbol \phi$, subject to the restriction \eqref{eq:identifiability}. Any standard statistical software may be used for this purpose; in this work, we use the $\mathtt{R}$ software for computations. 

The integral involved in \eqref{eq:lik-marg} can be evaluated numerically, by using any standard technique such as the quadrature methods. Here, we use the function $\mathtt{integrate()}$ in $\mathtt{R}$ software for the numerical evaluation of the integral, which is a quadrature-based technique. For the numerical optimization of \eqref{eq:lik-marg}, the $\mathtt{optim()}$ function of the $\mathtt{optim}$ package is used. 

\subsection{Alternative method based on Monte Carlo sampling} \label{subsec:fitting-MC}
It is possible to exploit the relationship between the gamma and the GG distribution to arrive at a simplified expression for the integral 
\begin{equation}
\int_{0}^{\infty}e^{-\frac{\eta_k(\omega(t_{ij};b_k))^2(z_k(t_{ij})-\lambda_i\mu_k\omega(t_{ij};b_k))^2}{2\lambda_i^2\mu_k^2(\omega(t_{ij};b_k))^2(z_k(t_{ij}))}}f_{gg}(\lambda_i;q,\sigma,\alpha)d\lambda_i \label{eq:lik-int}
\end{equation}
present in \eqref{eq:lik-marg}. Thus, using the simplified integral, a Monte Carlo sampling-based estimation method can be developed as follows. 

\subsubsection{Case: $q \neq 0$}
Define $u=(\alpha \lambda)^{\frac{q}{\sigma}}$. Using this transformation, it can be shown that  
\[
f_{gg}(\lambda;q,\sigma,\alpha) = f_g(u;q),
\]
where $f_g(u;q)=\frac{(q^{-2})^{(q^{-2})}}{\Gamma(q^{-2})} u^{(q^{-2}-1)}e^{-u(q^{-2})}$, which is the PDF of a gamma distribution; that is, $U=(\alpha \lambda)^{\frac{q}{\sigma}} \sim gamma \big(q^{-2},\frac{1}{q^{-2}} \big)$. 

Therefore, the integral in \eqref{eq:lik-int}, by using the transformation $u=(\alpha \lambda)^{\frac{q}{\sigma}}$, can be expressed as 
\begin{equation}
\begin{split}
    \int_{0}^{\infty}e^{-\frac{\eta_k(z_k(t_{ij})-\lambda_i\mu_k\omega(t_{ij};b_k))^2}{2\lambda_i^2\mu_k^2(z_k(t_{ij}))}}f_{gg}(\lambda_i;q,\sigma,\alpha)d\lambda_i & \nonumber \\
    &=\int_{0}^{\infty}e^{-\frac{\eta_k(z_k(t_{ij})-(\frac{u_i^{\frac{\sigma}{q}}}{\alpha})\mu_k\omega(t_{ij};b_k))^2}{2\mu_k^2(z_k(t_{ij}))(\frac{u_i^{\frac{2\sigma}{q}}}{\alpha^2})}}f_{g}(u_i;q)du_i,
\end{split}
\end{equation}
and consequently, the marginal likelihood in \eqref{eq:lik-marg} can be expressed as 
\begin{equation}
L_{marginal}(\boldsymbol \theta, \boldsymbol \phi|Data) =  \prod_{i=1}^n\prod_{k=1}^2\prod_{j=1}^{m_i}\sqrt{\frac{\eta_k}{2\pi (z_k(t_{ij}))^3}} \times \omega(t_{ij};b_k) \times \mathds{E}(h(U_i;\boldsymbol \theta, \boldsymbol \phi)), \label{eq:lik-q-neq-0}
\end{equation}
where
\[
h(U_i;\boldsymbol \theta, \boldsymbol \phi) = -\frac{\eta_k(z_k(t_{ij})-(\frac{u_i^{\frac{\sigma}{q}}}{\alpha})\mu_k\omega(t_{ij};b_k))^2}{2\mu_k^2(z_k(t_{ij}))(\frac{u_i^{\frac{2\sigma}{q}}}{\alpha^2})},
\]
\[
\mathds{E}(h(U_i;\boldsymbol \theta, \boldsymbol \phi)) = \int_{0}^{\infty}e^{-\frac{\eta_k(z_k(t_{ij})-(\frac{u_i^{\frac{\sigma}{q}}}{\alpha})\mu_k\omega(t_{ij};b_k))^2}{2\mu_k^2(z_k(t_{ij}))(\frac{u_i^{\frac{2\sigma}{q}}}{\alpha^2})}}f_{g}(u_i;q)du_i. 
\]
For evaluating \eqref{eq:lik-q-neq-0}, the expectation $\mathds{E}(h(U_i;\boldsymbol \theta, \boldsymbol \phi))$ can be approximated by 
\begin{equation} 
\mathds{E}(h(U_i;\boldsymbol \theta, \boldsymbol \phi)) \approx \frac{1}{R} \sum_{r=1}^{R} exp \Bigg\{ -\frac{\eta_k(z_k(t_{ij})-(\frac{u_{ir}^{\frac{\sigma}{q}}}{\alpha})\mu_k\omega(t_{ij};b_k))^2}{2\mu_k^2(z_k(t_{ij}))(\frac{u_{ir}^{\frac{2\sigma}{q}}}{\alpha^2})} \Bigg\}, \label{eq:exp-q-neq-0}
\end{equation}
by generating random samples $u_1,...,u_R$ from $gamma \big(q^{-2},\frac{1}{q^{-2}} \big)$. We recommend using a large sample from $gamma \big(q^{-2},\frac{1}{q^{-2}} \big)$ for this approximation, such as $R>10^5$.

\subsubsection{Case: $q \rightarrow 0$}
Note that when $q \longrightarrow 0$, $\alpha \longrightarrow 1$, the marginal likelihood function is 
$$L_{marginal}(\boldsymbol \theta, \boldsymbol \phi|Data)
=\prod_{i=1}^n\prod_{k=1}^2\prod_{j=1}^{m_i}\sqrt{\frac{\eta_k}{2\pi (z_k(t_{ij}))^3}} \times \omega(t_{ij};b_k)\\ \times \int_{0}^{\infty}e^{-\frac{\eta_k(z_k(t_{ij})-\lambda_i\mu_k\omega(t_{ij};b_k))^2}{2\lambda_i^2\mu_k^2(z_k(t_{ij}))}}f_{gg}(\lambda_i;\sigma)d\lambda_i. \label{eq:mar-lik-q=0}
$$

In this case, considering the transformation $u=(\lambda)^{\frac{1}{\sigma}}$, it can be shown that 
$$ 
f_{gg}(\lambda;\sigma) = f_{ln}(u),
$$
where $f_{ln}(u)= \frac{1}{\sqrt{2\pi}}e^{\frac{-(log( u))^2}{2 }}$ is the PDF of the standard lognormal distribution; that is, $U_i=(\lambda_i)^{\frac{1}{\sigma}} \sim Lognormal(0,1)$. 

Therefore, by using this transformation, the involved integral can be expressed as 

$$
\int_{0}^{\infty}e^{-\frac{\eta_k(z_k(t_{ij})-\lambda_i\mu_k\omega(t_{ij};b_k))^2}{2\lambda_i^2\mu_k^2(z_k(t_{ij}))}}f_{gg}(\lambda_i;q,\sigma,\alpha)d\lambda_i = \int_{0}^{\infty}e^{-\frac{\eta_k(z_k(t_{ij})-({u_i^{\sigma}})\mu_k\omega(t_{ij};b_k))^2}{2\mu_k^2(z_k(t_{ij}))({u_i^{2\sigma}})}}f_{g}(u_i;q)du_i.
$$


Consequently, the marginal likelihood in this case can be approximated as 
$$
L_{marginal}(\boldsymbol \theta,\boldsymbol \phi|Data)
 \approx \prod_{i=1}^n\prod_{k=1}^2\prod_{j=1}^{m_i}\sqrt{\frac{\eta_k}{2\pi (z_k(t_{ij}))^3}} \times \omega(t_{ij};b_k) \times \mathds{E}(h(U_i;\boldsymbol \theta,\boldsymbol \phi)),
$$
where
\begin{equation}
\mathds{E}(h(U_i;\boldsymbol \theta,\boldsymbol \phi)) \approx \frac{1}{R} \sum_{r=1}^{R} exp \Bigg\{ -\frac{\eta_k(z_k(t_{ij})-({u_{ir}^{\sigma}})\mu_k\omega(t_{ij};b_k))^2}{2\mu_k^2(z_k(t_{ij}))({u_{ir}^{2\sigma}})} \Bigg\}.  \label{eq:approx-lik-exp-q=0}
\end{equation}
The expectation in \eqref{eq:approx-lik-exp-q=0} can be evaluated by generating $R$ random samples $u_1,...,u_R$ from the standard lognormal distribution; recommended value of $R$ is $R>10^5$.

\section{Empirical Study} \label{sec:simulations}
The main aim of the Monte Carlo simulation study carried out here is to examine the efficiency of the fitting methods for the proposed model. Towards this, based on simulated bivariate degradation data, the model parameters are estimated, and the bias and root mean squared error (RMSE) of the estimates are calculated against their true values (which were used for simulating the bivariate degradation data). Evidently, smaller values of bias and RMSE indicate that the fitting methods work well in the setting considered.  

For simulations, we chose the number of units and degradation measurement times as $n = 50, 100$ and $m = 20, 40$. For the monotonic function to transform the time scale, we took $\Lambda(t;b_k)=t^{b_k}$, for $k=1,2$. For simulating marginal degradation measurements corresponding to the two PCs, the parameters of the IG process are taken as $\boldsymbol \theta=(\mu_1,\mu_2,\eta_1,\eta_2,b_1,b_2)$ $=(1.50, 1.00, 0.50, 1.00, 1.20, 0.80)$. For the shared frailty, we use the GG distribution with different choices of parameters, to ensure that we cover all the special members of the family, as given in Table \ref{tab:GG}.  The results of the Monte Carlo simulation study are presented in Tables \ref{tab:n50-m20} - \ref{tab:n100-m40}. The bias and RMSE values for all the models are reasonable, although the RMSE for $\mu_1$ and $\mu_2$ seem to be higher than the rest of the parameters, indicating some additional variability in estimating these two parameters.   
\begin{table}
\caption{Estimates of model parameters of the IG process and GG frailty based on bivariate degradation data; $n = 50$, $m=20$}

\begin{center}
	\begin{tabular}{c c c c c c c c c c} 
		\hline
		Parameter & $b_1$ & $\eta_1$ & $\mu_1$ & $b_2$ & $\eta_2$ & $\mu_2$ & $\sigma$ & $q$ & $\alpha$ \\ [0.5ex] 
		\hline
		\multicolumn{10}{c}{Gamma Frailty} \\
		\hline
		True value &  1.200 & 0.500 & 1.500 & 0.800 & 1.000 & 1.000 & 0.707 & 0.707 & 1.000
		\\ 
		Bias &  -0.002 & 0.002 &-0.005 &-0.002 & 0.005 & 0.024 &-0.011 &-0.045  &0.087
		\\
		RMSE &  0.021 &0.046 &0.464 &0.023& 0.075& 0.361 &0.265& 0.394 &0.292
		
		\\
		\hline
	\end{tabular}

\begin{tabular}{c c c c c c c c c c} 
	\hline
	Parameter & $b_1$ & $\eta_1$ & $\mu_1$ & $b_2$ & $\eta_2$ & $\mu_2$ & $\sigma$ & $q$ & $\alpha$ \\  
	\hline
	\multicolumn{10}{c}{Weibull Frailty} \\
	\hline
	True value &  1.200 & 0.500 & 1.500 & 0.800 & 1.000 & 1.000 & 0.697 & 1.000 & 0.908
	\\ 
	Bias & 0.003 & 0.009 & 0.032 & 0.000 & 0.013 & 0.027 &-0.023 &-0.019 & 0.055
	\\
	RMSE &  0.023 &0.047 &0.434 &0.023 &0.079 &0.323 &0.250 &0.306 &0.190
	\\
	\hline
\end{tabular}

\begin{tabular}{c c c c c c c c c c} 
\hline
 Parameter & $b_1$ & $\eta_1$ & $\mu_1$ & $b_2$ & $\eta_2$ & $\mu_2$ & $\sigma$ & $q$ & $\alpha$ \\  
 \hline
 \multicolumn{10}{c}{Lognormal frailty} \\
  \hline
 True value & 1.200& 0.500& 1.500 &0.800 &1.000& 1.000 &0.637 &0.000 &1.225
 \\ 
 Bias & 0.001 & 0.005 &-0.146 &-0.003& -0.002 &-0.041& -0.120 &-0.006& -0.021
 \\
 RMSE &  0.023 &0.048& 0.609& 0.023& 0.080 &0.287& 0.276 &0.147& 0.235
 \\
 \hline
 \end{tabular}  

\begin{tabular}{c c c c c c c c c c} 
	\hline
	Parameter & $b_1$ & $\eta_1$ & $\mu_1$ & $b_2$ & $\eta_2$ & $\mu_2$ & $\sigma$ & $q$ & $\alpha$ \\  
	\hline
	\multicolumn{10}{c}{Generalised Gamma Frailty} \\
	\hline
	True value &  1.200 & 0.500 & 1.500 & 0.800 & 1.000 & 1.000 & 0.642 & 1.500 & 0.775
	\\ 
	Bias &  0.001 & 0.007 & 0.140 & 0.003 & -0.001 & 0.017 & -0.127 & -0.022 & 0.042 \\
	RMSE &  0.024 & 0.053 & 0.561 & 0.021 & 0.078 & 0.293 & 0.226 & 0.421 & 0.097
	\\
	\hline
\end{tabular}

\label{tab:n50-m20}
\end{center}
\end{table}

\begin{table}
\caption{Estimates of model parameters of the IG process and GG frailty based on bivariate degradation data; $n = 50$, $m=40$}
\begin{center}
	
\begin{tabular}{c c c c c c c c c c} 
	\hline
	Parameter & $b_1$ & $\eta_1$ & $\mu_1$ & $b_2$ & $\eta_2$ & $\mu_2$ & $\sigma$ & $q$ & $\alpha$ \\ [0.5ex] 
	\hline
	\multicolumn{10}{c}{Gamma Frailty} \\
	\hline
	True value &  1.200 & 0.500 & 1.500 & 0.800 & 1.000 & 1.000 & 0.707 & 0.707 & 1.000
	\\ 
	Bias &  -0.002 &-0.002 & 0.031 &-0.001 & 0.002  &0.085 &-0.007& -0.021 & 0.066
	
	\\
	RMSE &  0.018 &0.037 &0.421& 0.015 &0.050 &0.339 &0.253 &0.387 &0.242
	
	\\
	\hline
\end{tabular}

\begin{tabular}{c c c c c c c c c c} 
	\hline
	Parameter & $b_1$ & $\eta_1$ & $\mu_1$ & $b_2$ & $\eta_2$ & $\mu_2$ & $\sigma$ & $q$ & $\alpha$ \\ [0.5ex] 
	\hline
	\multicolumn{10}{c}{Weibull Frailty} \\
	\hline
	True value &  1.200 & 0.500 & 1.500 & 0.800 & 1.000 & 1.000 & 0.697 & 1.000 & 0.908\\ 
	Bias &  -0.001 & 0.000 &-0.035 & 0.002 & 0.006 & 0.047 &-0.038 &-0.022 & 0.051\\
	RMSE &  0.016 &0.032 &0.461 &0.016 &0.058 &0.283 &0.236 &0.337 &0.172\\
	\hline
\end{tabular}

\begin{tabular}{c c c c c c c c c c} 
\hline
 Parameter & $b_1$ & $\eta_1$ & $\mu_1$ & $b_2$ & $\eta_2$ & $\mu_2$ & $\sigma$ & $q$ & $\alpha$ \\ 
 \hline
 \multicolumn{10}{c}{Lognormal Frailty} \\
  \hline
 True value & 1.200& 0.500& 1.500 &0.800 &1.000& 1.000 &0.637 &0.000 &1.225\\ 
 Bias & -0.001 &-0.003 &-0.056 &-0.003& -0.006& -0.001 &-0.101 & 0.003 &-0.017\\
 RMSE &  0.018 &0.034& 0.508& 0.014& 0.048 &0.281 &0.266 &0.126 &0.209\\
\hline
\end{tabular}  

\begin{tabular}{c c c c c c c c c c} 
	\hline
	Parameter & $b_1$ & $\eta_1$ & $\mu_1$ & $b_2$ & $\eta_2$ & $\mu_2$ & $\sigma$ & $q$ & $\alpha$ \\  
	\hline
	\multicolumn{10}{c}{Generalised Gamma Frailty} \\
	\hline
	True value &  1.200 & 0.500 & 1.500 & 0.800 & 1.000 & 1.000 & 0.642 & 1.500 & 0.775\\ 
	Bias &  0.000 & 0.002 & 0.082 &-0.002 &-0.007 & 0.014& -0.075 &-0.037  &0.033 \\
	RMSE & 0.017& 0.037& 0.420 &0.018 &0.071& 0.303 &0.177 &0.357 &0.083\\
	\hline
\end{tabular}
\label{tab:n50-m40}
\end{center}
\end{table}

\begin{table}
\caption{Estimates of model parameters of the IG process and GG frailty based on bivariate degradation data; $n = 100$, $m = 20$}
\begin{center}
	
\begin{tabular}{c c c c c c c c c c} 
	\hline
	Parameter & $b_1$ & $\eta_1$ & $\mu_1$ & $b_2$ & $\eta_2$ & $\mu_2$ & $\sigma$ & $q$ & $\alpha$ \\ [0.5ex] 
	\hline
	\multicolumn{10}{c}{Gamma Frailty} \\
	\hline
	True value &  1.200 & 0.500 & 1.500 & 0.800 & 1.000 & 1.000 & 0.707 & 0.707 & 1.000
	\\ 
	Bias &  0.002 & 0.006 & 0.082 &-0.001 & 0.001&  0.032 &-0.045 &-0.009&  0.046
	
	\\
	RMSE & 0.015& 0.032 &0.482 &0.017 &0.055& 0.272& 0.228 &0.429 &0.225
	
	\\
	\hline
\end{tabular}

\begin{tabular}{c c c c c c c c c c} 
	\hline
	Parameter & $b_1$ & $\eta_1$ & $\mu_1$ & $b_2$ & $\eta_2$ & $\mu_2$ & $\sigma$ & $q$ & $\alpha$ \\  
	\hline
	\multicolumn{10}{c}{Weibull Frailty} \\
	\hline
	True value &  1.200 & 0.500 & 1.500 & 0.800 & 1.000 & 1.000 & 0.697 & 1.000 & 0.908\\ 
	Bias & 0.000 & 0.002 & 0.095 & 0.000 & 0.002 & 0.007 &-0.012 &-0.086 & 0.051\\
	RMSE & 0.016 &0.036 &0.423 &0.017 &0.057 &0.217 &0.174 &0.289 &0.126\\
	\hline
\end{tabular}

\begin{tabular}{c c c c c c c c c c} 
\hline
 Parameter & $b_1$ & $\eta_1$ & $\mu_1$ & $b_2$ & $\eta_2$ & $\mu_2$ & $\sigma$ & $q$ & $\alpha$ \\ 
 \hline
 \multicolumn{10}{c}{Lognormal Frailty} \\
  \hline
 True value & 1.200& 0.500& 1.500 &0.800 &1.000& 1.000 &0.637 &0.000 &1.225\\ 
 Bias & 0.001 & 0.004& -0.010 & 0.000 & 0.006& -0.003& -0.092&  0.025& -0.022\\
 RMSE &  0.017 &0.035 &0.495 &0.014 &0.052 &0.240 &0.257 &0.136 &0.206\\
 \hline
\end{tabular}  

\begin{tabular}{c c c c c c c c c c} 
	\hline
	Parameter & $b_1$ & $\eta_1$ & $\mu_1$ & $b_2$ & $\eta_2$ & $\mu_2$ & $\sigma$ & $q$ & $\alpha$ \\  
	\hline
	\multicolumn{10}{c}{Generalised Gamma Frailty} \\
	\hline
	True value &  1.200 & 0.500 & 1.500 & 0.800 & 1.000 & 1.000 & 0.642 & 1.500 & 0.775\\ 
	Bias &  0.001 & 0.000 & 0.126 &0.003& -0.018  &0.041 &-0.117& -0.054  &0.049 \\
	RMSE & 0.018 &0.038 &0.452 &0.018 &0.078 &0.235 &0.213 &0.451 &0.108\\
	\hline
\end{tabular}
\label{tab:n100-m20}
\end{center}
\end{table}

\begin{table}
\caption{Estimates of model parameters of the IG process and GG frailty based on bivariate degradation data; $n = 100$, $m = 40$}
\begin{center}
	
\begin{tabular}{c c c c c c c c c c} 
	\hline
	Parameter & $b_1$ & $\eta_1$ & $\mu_1$ & $b_2$ & $\eta_2$ & $\mu_2$ & $\sigma$ & $q$ & $\alpha$ \\ [0.5ex] 
	\hline
	\multicolumn{10}{c}{Gamma Frailty} \\
	\hline
	True value &  1.200 & 0.500 & 1.500 & 0.800 & 1.000 & 1.000 & 0.707 & 0.707 & 1.000
	\\ 
	Bias &  0.001 & 0.002 & 0.107& -0.001& -0.002 & 0.037& -0.020 &-0.042 & 0.059
	\\
	RMSE & 0.012 &0.024 &0.491& 0.011& 0.037 &0.242 &0.218 &0.406& 0.203
	\\
	\hline
\end{tabular}

\begin{tabular}{c c c c c c c c c c} 
 \hline
 Parameter & $b_1$ & $\eta_1$ & $\mu_1$ & $b_2$ & $\eta_2$ & $\mu_2$ & $\sigma$ & $q$ & $\alpha$ \\ 
\hline
 \multicolumn{10}{c}{Weibull Frailty} \\
  \hline
 True value &  1.200 & 0.500 & 1.500 & 0.800 & 1.000 & 1.000 & 0.697 & 1.000 & 0.908\\ 
 Bias & 0.001 & 0.003 & 0.009 &-0.001 &-0.004 & 0.074 &-0.032 &-0.008 & 0.022\\
 RMSE & 0.011 &0.023 &0.342 &0.011 &0.037 &0.236 &0.173 &0.263 &0.105\\
 \hline
\end{tabular}

\begin{tabular}{c c c c c c c c c c} 
\hline
 Parameter & $b_1$ & $\eta_1$ & $\mu_1$ & $b_2$ & $\eta_2$ & $\mu_2$ & $\sigma$ & $q$ & $\alpha$ \\  
 \hline
 \multicolumn{10}{c}{Lognormal Frailty} \\
  \hline
 True value & 1.200& 0.500& 1.500 &0.800 &1.000& 1.000 &0.637 &0.000 &1.225\\ 
 Bias &0.000 & 0.002 &-0.068& -0.002 &-0.006 &-0.037 &-0.048 & 0.007 & 0.004\\
 RMSE &  0.012 &0.024& 0.456& 0.012 &0.042 &0.236& 0.228& 0.128 &0.184\\
 \hline
\end{tabular}  

\begin{tabular}{c c c c c c c c c c} 
	\hline
	Parameter & $b_1$ & $\eta_1$ & $\mu_1$ & $b_2$ & $\eta_2$ & $\mu_2$ & $\sigma$ & $q$ & $\alpha$ \\ 
	\hline
	\multicolumn{10}{c}{Generalized Gamma Frailty} \\
	\hline
	True value &  1.200 & 0.500 & 1.500 & 0.800 & 1.000 & 1.000 & 0.642 & 1.500 & 0.775\\ 
	Bias &  0.001 &-0.001 & 0.182 & 0.002 &-0.010 & 0.003 &-0.086& -0.118  &0.048 \\
	RMSE & 0.011 & 0.025 & 0.449 & 0.013 & 0.065 & 0.205 & 0.173 & 0.340 & 0.097\\
	\hline
\end{tabular}

\label{tab:n100-m40}
\end{center}
\end{table}

\section{Case Study: Fatigue Crack Data} \label{sec:real-data}
This case study demonstrates the practical process of model fitting and model selection for bivariate degradation data. The fatigue crack data, originally reported in Bogdanoff and Kozin~\cite{Bogdanoff-book}, was presented and analyzed by Meeker et al.~\cite{Meeker-book}, and Lu and Meeker~\cite{Lu-Meeker}. In Pan and Balakrishnan~\cite{Pan-Bala}, a sample of this fatigue crack was analyzed in the context of degradation modeling of products with multiple PCs based on gamma processes. Recently, Barui et al.~\cite{Barui} analyzed this data for bivariate degradation.   

The sample of the fatigue crack data analyzed by Pan and Balakrishnan~\cite{Pan-Bala} has information on the development of two fatigue cracks - Crack A and Crack B, on 10 units. The measurements are taken at the same measurement times for the two cracks, in the unit of million cycles. Each of the units is then measured at 10 measurement times, which are the same for all units: 0.00 to 0.09 million cycles with an interval of 0.01 million cycles.

\begin{table}
\tiny
\caption{Copula functions used for model fitting}
\begin{center}
\begin{tabular}{|c|c|} 
 \hline
 Name of Copula & Copula Function: $c(u_1, u_2)$   \\ 
 \hline
 Gaussian & $\frac{1}{\sqrt{1-\rho^2}}\exp{(-\frac{\rho^2 (\Phi^{-1}(u_1)^2 +\Phi^{-1}(u_2)^2)-2\rho \Phi^{-1}(u_1)\Phi^{-1}(u_2)}{2(1-\rho^2)})}$  \\ 
 Frank  &  $\frac{\delta (1-e^{-\delta}) e^{-\delta(u_1+u_2)}}{[(1-e^{-\delta})-(1-e^{-\delta u_1})(1-e^{-\delta u_2})]}$\\ 
 Clayton  & $(1+\delta)(u_1 u_2)^{-\delta-1} (u_1^{-\delta} + u_2^{-\delta} -1 )^{-\frac{1}{\delta}-2}$ \\
 Gumbel & $
\exp(-[(-\ln u_1)^{\delta}+(-\ln u_2)^{\delta}]^{\frac{1}{\delta}}) \frac{1}{u_1 u_2} [(-\ln u_1)^{\delta}+(-\ln u_2)^{\delta}]^{-2+\frac{2}{\delta}}] 
\times (\ln u_1 \ln u_2)^{\delta-1}(1+(\delta-1)[(-\ln u_1)^{\delta}+(-\ln u_2)^{\delta}]^{\frac{-1}{\delta}})$ \\
FGM & $u_1u_2 + \delta \cdot u_1(1-u_1)u_2(1-u_2)$ \\
 Independent & 1 \\
\hline
\end{tabular} \label{tab:copulas}
\footnotesize{$\rho$ is the correlation between two certain random variables; $dt(, \gamma)$ is the probability density of univariate $t_\gamma$ distribution}
\end{center}
\end{table}

We fit the proposed IG-GG model to this fatigue crack data. For comparative purposes, we also fit various copula-based models with IG processes for the marginal degradations to the fatigue crack data. For the copula-based models, while the marginal degradation processes are modeled by the IG processes, we use various copula functions for dependence modeling. The copulas used here for comparative purposes are given in Table \ref{tab:copulas}. Different copulas have different ranges for the copula parameter; see Grober~\cite{Grober} for details. 

Parameters corresponding to all the fitted models are estimated based on the fatigue-crack data. Correspondingly, the maximized values of the log-likelihood functions and the model selection criteria (e.g., AIC, BIC etc.) are calculated for all the fitted models. Finally, the model with the lowest AIC (or BIC) is chosen as the best model for the fatigue-crack data. Estimated parameters of the different models are presented in Table \ref{tab:fc-par-est}. The maximized log-likelihood, AIC, and BIC values are presented in Table \ref{tab:fc-aic}.  

\begin{table}
	\caption{Estimated parameter values for different models fitted to the fatigue-crack data}
	\begin{center}
    \resizebox{\columnwidth}{!}{
		\begin{tabular}{|c|c|c|c|c|c|c|c|c|c|c|} 
			\hline
			Model & $b_1$ & $\eta_1$ & $\mu_1$ & $b_2$ & $\eta_2$ & $\mu_2$ & $\sigma$ & $q$ & $\alpha$ & $\delta$ \\ 
			\hline
    IG - Exp &     1.082 &  5227.638 &    10.104 &     1.687 &  7617.269 & 71043.296 &     1.000 &     1.000 &     1.000 &        -    \\

   IG - Gam  &     1.074 &  5742.666 &     7.102 &     1.613 &  4987.721 & 68044.612 &     0.277 &     0.277 &     1.000 &          -  \\

   IG - Wei  &     1.180 & 12888.390 &     9.183 &     1.582 &  3973.051 & 69859.792 &     0.072 &     1.000 &     0.963 &        -    \\

     IG - LN &     1.102 &  5890.927 &     7.584 &     1.682 &  7620.945 & 71036.838 &     0.026 &     0.000 &     1.000 &       -     \\

    IG - GG  &     1.184 & 19603.227 &     9.161 &     1.587 &  4240.866 & 69861.539 &     0.097 &     0.318 &     0.989 &         -   \\

Ind. Copula &     1.153 &  1145.505 &  4212.953 &     1.631 &  5587.272 & 49334.157 &     -       &    -        &    -        &     0.000 \\

  Clayton  &     1.311 & 19603.225 &    11.387 &     1.859 &  4240.875 & 69861.539 &     -       &    -        &    -        &     0.241 \\

     Frank &     1.023 &  1145.489 &  4212.953 &     1.348 &  5587.287 & 49334.157 &        -    &     -       &   -         &     7.386 \\

    Gumbel &     1.195 &  5227.634 &    12.873 &     1.613 &  7617.272 & 71043.296 &      -      &     -       &    -        &     0.010 \\

  Gaussian &     1.076 &  1145.504 &  4212.953 &     1.412 &  5587.272 & 49334.157 &     -       &    -        &      -      &     0.816 \\

       FGM &     1.090 &  5227.634 &     7.375 &     1.620 &  7617.275 & 71043.296 &       -     &       -     &     -       &     0.010 \\
			\hline
		\end{tabular} \label{tab:fc-par-est}}
	\end{center}
\end{table}

\begin{table}
	\caption{Log-likelihood, AIC, and BIC values for different fitted models for the fatigue-crack data}
	\begin{center}
		\begin{tabular}{|c|c|c|c|} 
			\hline
			Model & Log-likelihood value & AIC value & BIC value \\ 
			\hline
              IG - Exp &   327.314 &  -642.627 & -640.811 \\

 IG - Gam  &   393.038 &  -772.077 & -769.959 \\

 IG - Wei  &   406.263 &  -798.526 & -796.408 \\

   IG - LN &   394.439 &  -774.877 & -772.759 \\

{\bf IG - GG}  &   {\bf 410.660} &  {\bf -805.320} & {\bf -802.899}\\

Ind. Copula &   291.475 &  -570.950 & -569.135 \\

  Clayton  &   341.670 &  -669.341 & -667.223 \\

     Frank &   400.691 &  -787.383 & -785.265 \\

    Gumbel &   397.294 &  -780.589 & -778.471 \\

  Gaussian &   366.348 &  -718.696 & -716.578 \\

       FGM &   409.292 &  -804.584 & -802.466 \\

			\hline
		\end{tabular} \label{tab:fc-aic}
	\end{center}
\end{table}


The IG process model with GG frailty is chosen as the most suitable model for the fatigue-crack data. The estimated parameters for this model, based on the fatigue-crack data, are as follows, with the corresponding standard errors in parentheses: $\widehat{b}_1 = 1.082 (0.015)$, $\widehat{\eta}_1 = 19603.227 (443.412)$, $\widehat{\mu}_1 = 9.161 (0.322)$, $\widehat{b}_2 = 1.587 (0.043)$, $\widehat{\eta}_2 = 4240.866 (651.750)$, $\widehat{\mu}_2 = 69861.539 (612.854)$, $\widehat{\sigma} = 0.097 (0.006)$, $\widehat{q} = 0.318 (2.336)$, $\widehat{\alpha} = 0.989$. These estimates can be used further to assess various reliability characteristics of the unit with two degrading components as a whole. For example, if the interest is to estimate system reliability at a mission time, it would be first necessary to develop the reliability function for the system in this setting of the IG-GG model framework, and then to use the estimated model parameters of the IG-GG model to estimate the system reliability. This is discussed in the next section.

\section{Estimation of System Reliability} \label{sec:reliability}
Assume that the correlation coefficient between the components of the bivariate degradation increments $Z_1(t)$ and $Z_2(t)$ is $\rho$, which is a constant, not changing with measurement time $t$; obviously, $-1 < \rho < 1$. In this section, we develop an approximation for the reliability function of the product with two degrading components, following the results of Park and Padgett~\cite{Park-Padgett} and the work of Pan and Balakrishnan~\cite{Pan-Bala}.

Let $T_1$ and $T_2$ be the first passage times of IG processes $X_1$ and $X_2$, with threshold values $\tau_1$ and $\tau_2$, respectively. Assume that the measurement frequency is a constant for both processes, i.e., $\omega(t_j;b_k)=\omega(t; b_k)$, $k=1,2$. Also, recall that 
\[
Z_k(t_j) \sim IG(\lambda \mu_k \omega(t_j;b_k), \eta_k(\omega(t_j;b_k))^2), \quad k=1,2.
\]

Therefore, using the mean and variance of the IG process, $Z_k(t_j)$ can be normalized as 
\[
Y_k(t_j) = \frac{Z_k(t_j)-\lambda \mu_k \omega(t;b_k)}{\sqrt{\frac{(\lambda \mu_k)^3\omega(t;b_k)}{\eta_k}}}.
\] 
Here, $\boldsymbol Y = (Y_1(t_j), Y_2(t_j))$ are \textit{i.i.d.} vectors for different $t_j$, with 
\[
E[Y_1(t_j)] = E[Y_2(t_j)] = 0,
\]
\[
Var[Y_1(t_j)] = Var[Y_2(t_j)] = 1,
\]
\[
Corr(Y_1(t_j), Y_2(t_j)) = \rho. 
\]

It is easier to develop the expression for the reliability function first by considering the discrete versions of the first passage times and then to convert the expression for continuous first passage times, as done by Pan and Balakrishnan~\cite{Pan-Bala}. Therefore, we consider the discrete versions of $T_1$ and $T_2$, and denote them by $M_1$ and $M_2$, respectively. As $X_1$ and $X_2$ are both strictly increasing processes, the survival probability of the product after $m_1$ and $m_2$ measurements for the two PCs, respectively, conditional on the shared random effect $\lambda$, is given by
\begin{align}
	P[M_1 > m_1, M_2 > m_2 | \lambda] &= P[X_{1M_1} < \tau_1, X_{2M_2} <\tau_2]  \nonumber \\
	&= P\bigg[\sum_{j=1}^{m_1}Z_1(t_j)<\tau_1, \sum_{j=1}^{m_2}Z_2(t_j)<\tau_2\bigg] \nonumber \\
	&= P\bigg[\sum_{j=1}^{m_1}y_1(t_j)<\frac{\tau_1-\lambda \mu_1 m_1\omega(t;b_1)}{\sqrt{\frac{(\lambda \mu_1)^3\omega(t;b_1)}{\eta_1}}}, \sum_{j=1}^{m_2}y_2(t_j)<\frac{\tau_2-\lambda \mu_2 m_2\omega(t;b_2)}{\sqrt{\frac{(\lambda \mu_2)^3\omega(t;b_2)}{\eta_2}}}\bigg] \nonumber \\
	&= P\bigg[\frac{1}{\sqrt{m_1}}\sum_{j=1}^{m_1}y_1(t_j)<\frac{\tau_1-\lambda \mu_1 m_1\omega(t;b_1)}{\sqrt{\frac{(\lambda \mu_1)^3m_1\omega(t;b_1)}{\eta_1}}}, \frac{1}{\sqrt{m_2}}\sum_{j=1}^{m_2}y_2(t_j)<\frac{\tau_2-\lambda \mu_2 m_2\omega(t;b_2)}{\sqrt{\frac{(\lambda \mu_2)^3m_2\omega(t;b_2)}{\eta_2}}}\bigg]. \nonumber 
\end{align}
Now, by applying the bivariate central limit theorem~\cite{Hunter}, we obtain 
\begin{equation}
	P[M_1>m_1, M_2>m_2 | \lambda] \approx \Phi_2\left(\frac{\tau_1-\lambda \mu_1 m_1\omega(t;b_1)}{\sqrt{\frac{(\lambda \mu_1)^3m_1\omega(t;b_1)}{\eta_1}}}, \frac{\tau_2-\lambda \mu_2 m_2\omega(t;b_2)}{\sqrt{\frac{(\lambda \mu_2)^3m_2\omega(t;b_2)}{\eta_2}}}; \delta \rho \right), \label{eq:rel-int-1}
\end{equation}
where $\Phi_2(\cdot)$ is the cumulative distribution function of the bivariate normal distribution, and $\delta = min\left\{\sqrt{\frac{m_1}{m_2}},\sqrt{\frac{m_2}{m_1}}\right\}$. Using properties of bivariate normal distribution, $P[M_1>m_1, M_2>m_2]$ can be expressed as  
\begin{align}
	P[M_1>m_1, M_2>m_2 | \lambda] &\approx 1 - \Phi\left(\frac{\lambda \mu_1 m_1\omega(t;b_1) - \tau_1}{\sqrt{\frac{(\lambda \mu_1)^3m_1\omega(t;b_1)}{\eta_1}}}\right) - \Phi\left(\frac{\lambda \mu_2 m_2\omega(t;b_2)-\tau_2}{\sqrt{\frac{(\lambda \mu_2)^3m_2\omega(t;b_2)}{\eta_2}}}\right) \nonumber \\
	&+ \Phi_2\left(\frac{\lambda \mu_1 m_1\omega(t;b_1)-\tau_1}{\sqrt{\frac{(\lambda \mu_1)^3m_1\omega(t;b_1)}{\eta_1}}}, \frac{\lambda \mu_2 m_2\omega(t;b_2)-\tau_2}{\sqrt{\frac{(\lambda \mu_2)^3m_2\omega(t;b_2)}{\eta_2}}}; \delta \rho \right). \label{eq:rel-int-2}
\end{align}  

We may now switch to the continuous versions of the first passage times $T_1$ and $T_2$, and express the reliability function of the product, conditional of the shared frailty $\lambda$, as follows:
\begin{align}
	P[T_1>t_1, T_2>t_2 | \lambda] &\approx 1 - \Phi\left(\frac{\lambda \mu_1 \Lambda(t_1;b_1) - \tau_1}{\sqrt{\frac{(\lambda \mu_1)^3\Lambda(t_1;b_1)}{\eta_1}}}\right) - \Phi\left(\frac{\lambda \mu_2 \Lambda(t_2;b_2)-\tau_2}{\sqrt{\frac{(\lambda \mu_2)^3\Lambda(t_2;b_2)}{\eta_2}}}\right) \nonumber \\
	&+ \Phi_2\left(\frac{\lambda \mu_1 \Lambda(t_1;b_1)-\tau_1}{\sqrt{\frac{(\lambda \mu_1)^3\Lambda(t_1;b_1)}{\eta_1}}}, \frac{\lambda \mu_2 \Lambda(t_2;b_2)-\tau_2}{\sqrt{\frac{(\lambda \mu_2)^3\Lambda(t_2;b_2)}{\eta_2}}}; \delta \rho \right) \nonumber \\
	&= R(t_1, t_2 | \lambda). \label{eq:rel-int-3}
\end{align}  
If we consider the reliability of the unit at time $t_1 = t_2 = t$, we shall simply denote the reliability function in \eqref{eq:rel-int-3} by $R(t|\lambda)$. Finally, the unconditional or marginal reliability function of the product at time $t$ can be obtained as 
\begin{equation}
	R(t) = \int_{0}^{\infty}R(t|\lambda)f(\lambda)d\lambda. \label{eq:rel}
\end{equation} 
The reliability function in \eqref{eq:rel} can be calculated by using numerical integration quite easily. 


\section{Concluding Remarks} \label{sec:conclusion}
In this paper, we develop a frailty-based model for bivariate degradation data, by using the IG process as the marginal model for the two degrading PCs. For the frailty, we use the family of GG distributions that includes exponential, gamma, Weibull, and lognormal distributions as special cases. The IG-GG modeling framework has the primary and main advantage of being flexible in modeling bivariate degradation data. Along with modeling the individual degradation data corresponding to the two PCs by IG processes, it also captures the dependence between the degrading PCs suitably, without restricting the form of dependence to specific well-known frailty distributions such as gamma, Weibull, or lognormal. Due to the flexible form of the GG distribution, one can choose the most suitable frailty distribution among the GG family that provides the best fit to a given bivariate degradation data. Likelihood-based fitting methods that are convenient and simple to use are discussed, and the efficiency of the fitting methods are demonstrated through detailed simulation studies. For the real bivariate degradation data on fatigue cracks, we compare the proposed IG-GG model with many copula-based models which are commonly used for bivariate degradation data, and show that the IG-GG model provides the best fit. We also provide an expression for the reliability of an unit with two degrading PCs in this setting of the IG-GG model, and illustrate its computation based on the fatigue crack data.   

\section*{Conflict of Interest}
The authors do not have any conflict of interest to declare.


\bibliographystyle{plain}
\bibliography{papers}

\end{document}